\documentclass[preprint2]{aastex}

\usepackage{amsmath}

\shorttitle{National Astronomical Observatory of Japan}
\shortauthors{Iye}

\begin{document}


\title{National Astronomical Observatory of Japan}


\author{Masanori Iye\altaffilmark{1}}
\altaffiltext{1}{Optical and Infrared Astronomy Division, National Astronomical 
Observatory, Mitaka, Tokyo, 181-8588 Japan}



\begin{abstract}
National Astronomical Observatory is an inter-university institute serving as the national center for ground based astronomy offering observational facilities  covering the optical, infrared, radio wavelength domain.  NAOJ also has solar physics and geo-lunar science groups collaborating with JAXA for space missions and a theoretical group with computer simulation facilities.
The outline of NAOJ, its various unique facilities, and some highlights of recent science achievements are reviewed.  
\end{abstract}





\section{The Outline of NAOJ}

\begin{figure}[htbp]
\label{naojmap}
\begin{center}
\includegraphics[width=7cm,clip]{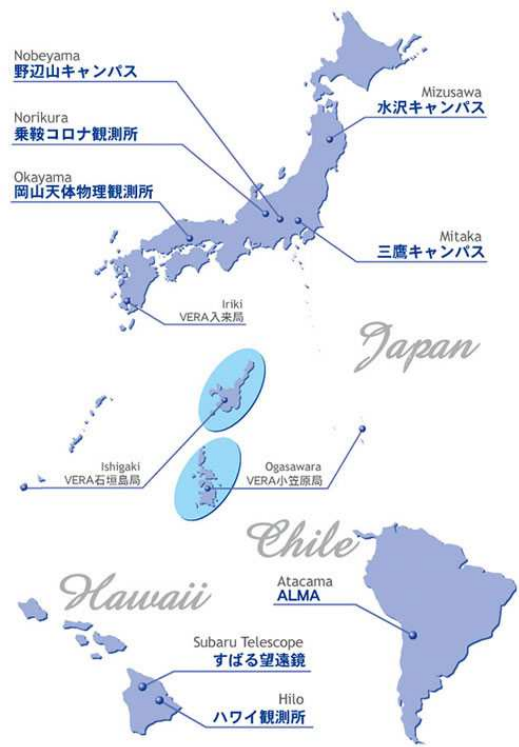}
\caption{NAOJ's five observatory campuses and three VERA stations. Also shown are Subaru Telescope facilities in Hawaii and ALMA facility under construction in Chile.
}
\end{center}
\end{figure}

The National Astronomical Observatory of Japan (NAOJ), as the national center of astronomical researches in Japan, deploys five observatories and three VERA stations in Japan, Subaru Telescope in Hawaii, and ALMA Observatory in Chile as shown in Fig.1.  NAOJ, belonging to the National Institutes for Natural Sciences (NINS) as one of the five inter-university institutes, offers its various top-level research facilities for researchers in the world.

Major facilities of NAOJ include 8.2m Subaru Telescope at Mauna Kea, Hawaii, 45m Radio Telescope and Radio Heliograph at Nobeyama, 1.8m telescope at Okayama Astrophysical Observatory, VERA interferometer for Galactic radio astrometry, solar facilities at Mitaka and Norikura, super computing facility and others.  In addition to these ground based facilities, NAOJ scientists have been core members for some of the ISAS space missions, Hinode, VSOP, and Kaguya.

The total number of NAOJ staff in 2007 amounts to 258, including 33 professors, 48 associate professors, and 82 research associates.  In addition, about 200 contract staffs are supporting the daily activities at each campus of NAOJ. A total of 72 graduate students and 34 post doctoral fellows are affiliated to NAOJ.

The annual total budget of NAOJ delivered from the Ministry of Education, Science, Culture and Sports, MEXT, for the fiscal year 2009 was 15.4 BY\!\!\!\!=, including 3.3 BY\!\!\!\!= for personnel expenses, 8.1 BY\!\!\!\!= for research expenses, and 3.5 BY\!\!\!\!= for facility development expenses.   In addition, 66 competitive research funds, amounting 0.6 BY\!\!\!\!= were acquired by applications of individual researchers in FY2008.

All the NAOJ facilities are offered for open use for scientists in Japan and some are offered for world wide community.  The selection process of successful proposals is based on the refereeing system for each facility and these processes are overseen by dedicated committees.   In the fiscal year 2007, a total of 354 proposals were approved and a total of 1300 researchers, including 200 foreigners, used NAOJ facilities for researches.

Observational and theoretical research achievements spanning the solar system, stars and clusters, interstellar matter, galaxies, active galactic nuclei, and cosmology were published in 376 refereed papers and 603 proceedings papers in English in FY2007.

NAOJ, like all national universities and inter-university institutes, was reformed to be a semi-governmental agency back in 2004, according to the structural reformation scheme of the government. The science achievements of NAOJ are highly respected and the government continues its financial supports, but this transformation in the academic system generated some turbulence.  NAOJ regards, however, its increased flexibility as a merit to plan for the better future taking the recommendations by external panel reviews.

NAOJ pays much effort also to promote public relations.  NAOJ opens its 4D2U dome theater, inaugurated in 2007, as a new facility to show original numerical simulation videos in three dimensions using stereographic technology.  NAOJ home page is one of the most popular site among the research institutes and universities in Japan. More than million accesses are observed every month, regularly.

\begin{figure}[htbp]
\label{Subaru}
\begin{center}
\includegraphics[width=8cm,clip]{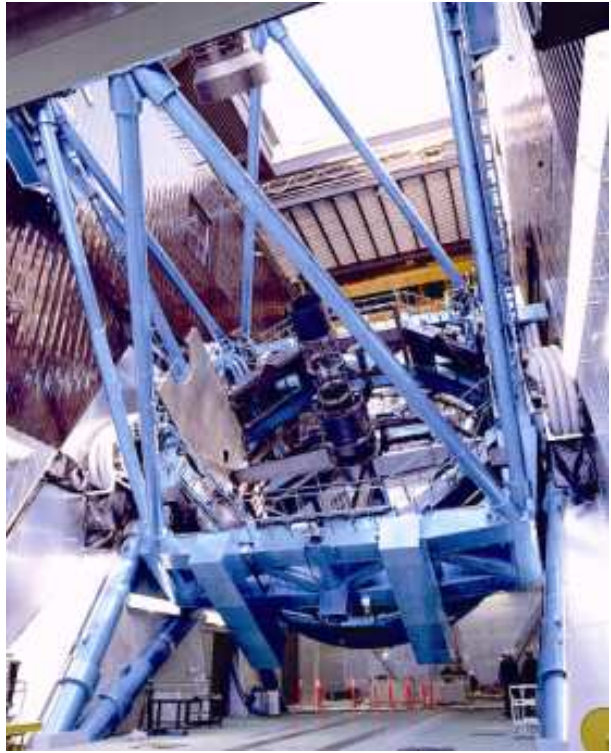}
\caption{8m Subaru Telescope has four foci and 8 open-use scientific instruments.
}
\end{center}
\end{figure}

\begin{figure}[htbp]
\label{Subaru}
\begin{center}
\includegraphics[width=7cm,clip]{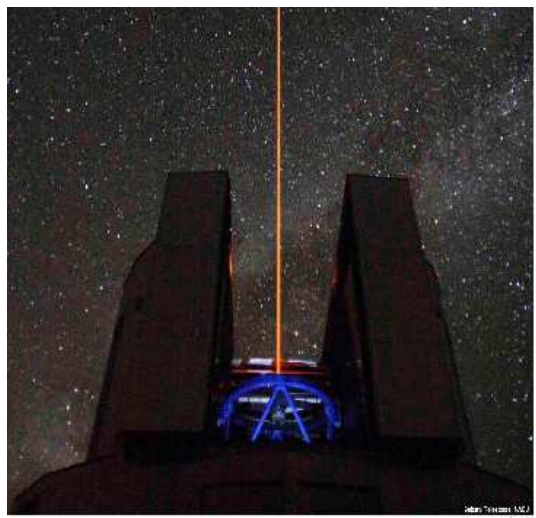}
\caption{New laser guide star adaptive optics system will enhance the spatial resolution of Subaru Telescope by an order of magnitude.
}
\end{center}
\end{figure}

\section{Subaru Telescope}

The 8.2m Subaru Telescope (Fig.2.), completed in 1999 atop the summit of 4200-meter Mauna Kea on the island of Hawai'i, provides astronomers of Japan and around the world with opportunities to carry out top-level observations in the optical to mid-infrared (0.3-30 $\mu$m) domain. Observational targets range widely, e.g., objects in our Solar system, putative planets around nearby stars, galaxies and clusters of galaxies, even to the furthest objects at the edge of the observable universe. In order to address this wide range of research interests, Subaru offers a suite of eight high performance cameras and spectrographs in the optical and infrared domain.

Continuous R\&D efforts to design and build new instruments are made possible by winning significant science budgets.  A new adaptive optics system, AO188, was constructed in 2006 to provide diffraction limited imaging capability in the near infrared bands with Strehl ration as high as 0.6. This system, together with a newly developed laser-guide star generating system (Fig.3.), which creates an artificial reference star in the upper atmosphere to operate the AO188 anywhere in the sky, will greatly enhance the Subaru capability from 2009. New instruments, HiCIAO, the successor to the current stellar coronagraph CIAO, to be used with AO188, Fiber Multi-Object Spectrograph, FMOS, with 400 fiber feed, and the HyperSuprimeCam, the successor of imager SuprimeCam, providing 10 times larger field coverage, will be on line shortly. 

NAOJ set up Extremely Large Telescope (ELT) project office in 2005 and hopes to establish an international partner ship to realize the Thirty Meter Telescope (TMT) project as a reality at Mauna Kea by around 2018.

\begin{figure}[htbp]
\label{IOK-1}
\begin{center}
\includegraphics[width=6cm,clip]{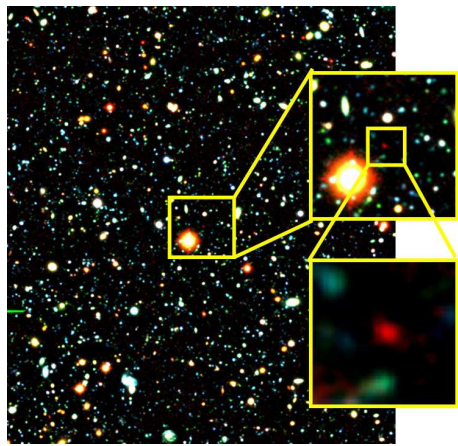}
\caption{The most distant galaxy, IOK-1, at 12.88 billion light years away.
}
\end{center}
\end{figure}

Subaru Telescope has been yielding many science results, discovery of the most distant galaxy at redshift 6.96 \citep{iye06} as shown in Fig.4, led to the new insight on the cosmic dawn of reionization of the universe.
It is noteworthy that nine from the top-ten most distant galaxies have been discovered by the Subaru Telescope as of March 2009.  Subaru also successfully mapped the distribution of dark matter through weak lensing analysis.

\begin{figure}[htbp]
\label{EMP-star}
\begin{center}
\includegraphics[width=7cm,clip]{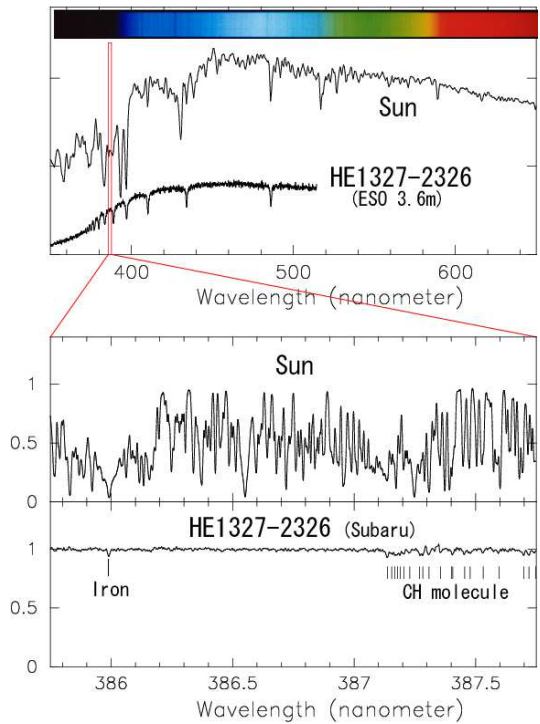}
\caption{High dispersion spectra of the Sun(above) and the most metal deficient star ever found, HE1327-2326(below). This star is likely to be the first generations of stars in the universe.
}
\end{center}
\end{figure}

Chemical abundance studies of old stars in the Galaxy are other areas, where Subaru made significant science achievements.  Fig.5 shows the most metal deficient star \citep{aoki06} discovered by Subaru High Dispersion Spectrograph, HDS.

\begin{figure}[htbp]
\label{yagi}
\begin{center}
\includegraphics[width=5cm,clip]{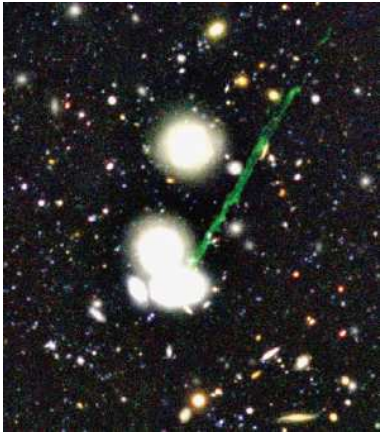}
\caption{Peculiar streak of HII gas found in the Coma Cluster
}
\end{center}
\end{figure}

Another new surprise brought by Subaru's Suprime-Cam camera is the discovery of  an unusual streak of ionized hydrogen gas associated with a galaxy 300 million light-years from Earth (Fig.6) \citep{yagi07}. The filament of gas is only 6 thousand light-years wide, yet extends 200,000 light-years, about the distance between the Milky Way Galaxy and its companion, the Large Magellanic Cloud. Finding such an extremely narrow and long ionized gas cloud is something quite unexpected.

\begin{figure}[htbp]
\label{hr8799}
\begin{center}
\includegraphics[width=5cm,clip]{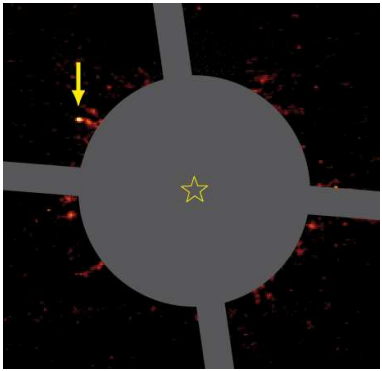}
\caption{Planet HR8799b as imaged by Subaru in 2002.
}
\end{center}
\end{figure}

Fig. 7 shows the image of the outermost planet HR 8799b as confirmed in the image taken by Subaru Telescope back in 2002.  The data were re-examined stimulated by the discovery paper of three planets around HR 8799 in 2008 \citep{maro08}. This measurement is used to the determination of the orbit of this planet.

\section{Atacama Large Millimeter/submillimeter Array(ALMA)}

ALMA is an international radio astronomy facility that is under construction on the Chilean plateau at an altitude of 5,000m above sea level. ALMA is a partnership between Japan, Europe and North America in cooperation with the Republic of Chile. NAOJ is leading the construction and operation of ALMA on behalf of the Japanese and Taiwanese science communities. ALMA consists of an array of sixty-eight 12-m antennas and twelve 7-m antennas extending over 18.5km span at maximum, allowing high spatial resolution equivalent to that of a telescope of the same size. NAOJ is building Atakama Compact Array as part of the ALMA project (Fig.8). The resolution of ALMA will be 10 times as high as that of the world's most sophisticated optical telescopes, such as Subaru. With such a high resolution one can recognize a coin from a distance of 400km. ALMA aims to explore the dark and invisible universe by detecting millimeter/submillimeter waves between radio wave and far-infrared spectral regions in order to figure out the mysteries on the birth of stars and planets, as well as the synthesis of organic molecules in the universe. ALMA is scheduled to start its full operation from 2012.

\begin{figure}[htbp]
\label{ACA}
\begin{center}
\includegraphics[width=8cm,clip]{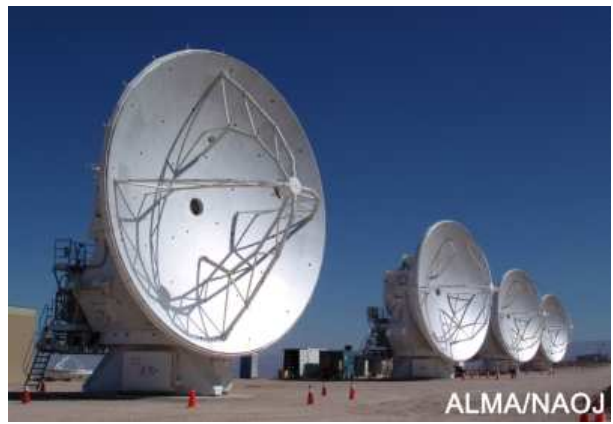}
\caption{Four 12m anntenas for Atacama Compact Array (ACA) as delivered to Chile.}
\end{center}
\end{figure}

\section{Nobeyama Radio Observatory}
The Nobeyama Radio Observatory (NRO) has two large radio telescopes, the Nobeyama 45m telescope and the Nobeyama Millimeter Array(NMA). The 45m telescope has been the largest radio telescope operated at millimeter wavelengths since its commissioning in 1982. The advantage of the 45m telescope is its high sensitivity. NMA consists of six 10m movable parabola antennas and provides high angular resolution, 10 times better than that of the 45m telescope. They have benn widely used for various observations of star-forming regions in our galaxy to distant galaxies, however NMA ceased its science operation since FY2008.

The discovery of the kinematical evidence of a massive black hole at the center of a galaxy NGC4258 by VLBA in NRAO/USA as shown in Fig.9 \citep{miyoshi} was based on the previous discovery of very high velocity water vapor masers by the 45m telescope. 

ASTE (Atacama Submillimeter Telescope Experiment) is the first 10-m submillimeter telescope in the southern hemisphere located 4800 m above sea level at the Atacama desert in Chile and operated by NRO/NAOJ and University of Tokyo in collaboration with the University of Chile etc. One of the important purposes of ASTE is exploring the southern sky at submillimeter wavelengths from 1 mm down to 0.3 mm (up to 950 GHz). ASTE equipped with 1.1 mm continuum camera performed the intensive surveys of distant dusty forming galaxies during 2007 to 2008 and discovered a cluster of extreme star-forming galaxies (Fig.10) which is located 11.5 billion light-years away from the Earth and is associated with a large scale structure of galaxies \citep{tamura09}.

\begin{figure}[tbph]
\label{miyoshi}
\begin{center}
\includegraphics[width=7cm,clip]{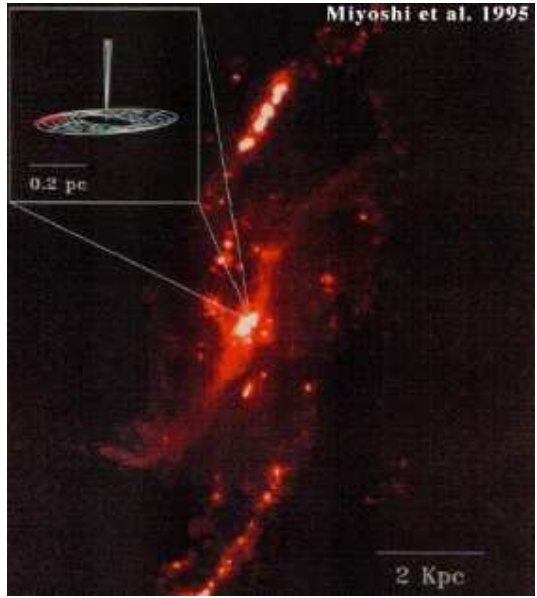}
\caption{Discovery of massive black hole at the center of NGC4258, proven by the Keplerian motion of water maser sources.}
\end{center}
\end{figure}

\begin{figure}[htbp]
\label{tamura09}
\begin{center}
\includegraphics[width=7cm,clip]{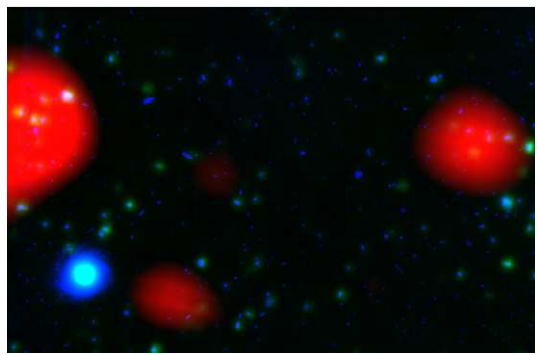}
\caption{Submillimeter galaxies (red) detected by ASTE superposed on the distribution of infrared sources (green) and optical sources (blue).
}
\end{center}
\end{figure}

\section{Mizusawa VERA Observatory}

VERA  (VLBI Exploration of Radio Astrometry) consists of four radio telescopes deployed over the Japanese archipelago to measure the parallax and proper motion of maser sources in our Galaxy using the Very Long Baseline Interferometry (VLBI) technique. VERA uses a two-beam observing system, which makes it possible to observe two celestial objects simultaneously. This enables the phase fluctuation caused by the atmospheric turbulence to be compensated for, and as a result, the positions of the radio sources can be determined very accurately. VERA is also able to determine the distance to objects with 10 micro-arc-second accuracy.  This is a kind of triangulation method based on the Earth's orbit. 

VERA demonstrated (cf. Fig.11) its high capability of precise astrometry by breaking the world record of measuring the largest distance by means of trigonometric parallax \citep{honma07} as well as by providing the best distance measurement for Orion KL, one of the most important Galactic objects.

\begin{figure}[htbp]
\label{vera}
\begin{center}
\includegraphics[width=7cm,clip]{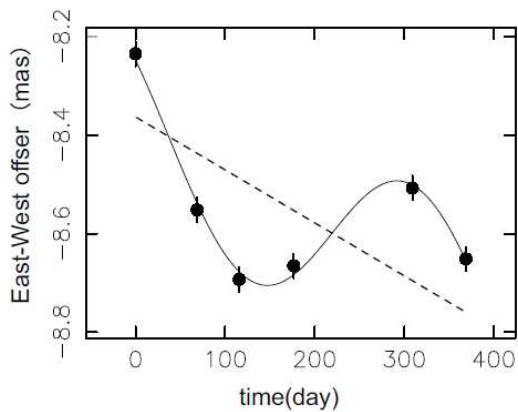}
\caption{Annual parallax of a water maser source in the star forming region S269 was measured by VERA for the first time to be $189\pm8$ micro arcsec.
}
\end{center}
\end{figure}

\section{Okayama Astrophysical Observatory}
 The 1.88m telescope at Okayama Astrophysical Observatory (OAO) \citep{yoshida05} was the largest optical infrared telescope and the heart of the modern astrophysics in Japan during 1960-1990s. The telescope has been used as a common use facility and more than 200 researchers are visiting at OAO. High dispersion optical spectroscopy of stars is the main research activity \citep{sada05}. One of the recent achievements is discovery of ten extra-solar planets around G-giant stars \citep{sato08}. Fig. 12 shows an example of the orbital modulation of radial velocity of such a star. There is a plan to construct a 3.8m telescope in the Okayama campus as a new facility of Kyoto University.

\begin{figure}[h]
\label{sato}
\begin{center}
\includegraphics[width=7cm,clip]{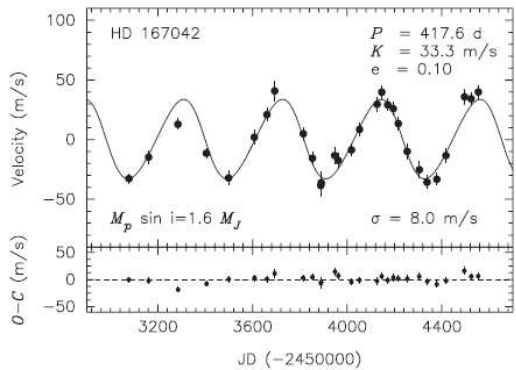}
\caption{Observed radial velocities of HD167042 (dots). The Keplerian orbital fit is shown by the solid line.
}
\end{center}
\end{figure}

\section{TAMA300 Gravitational Wave Detection Experiment}
The gravitational wave detection facility TAMA300, with a Fabry-Perot interferometer of 300m span to measure the time varying deformation of the space due to the gravitational wave, has been in contiguous operation since 1999. Sensitivity of the system has been much improved by the development of the new seismic attenuation system as is shown in Fig.13 \citep{arai07}.

\begin{figure}[htbp]
\label{tama300}
\begin{center}
\includegraphics[width=8cm,clip]{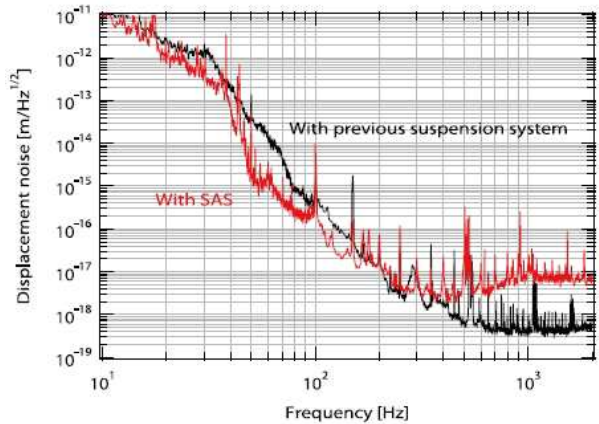}
\caption{Tama300 sensitivity with new seismic attenuation system (SAS) compared with that of old SAS.
}
\end{center}
\end{figure}

\section{Kaguya-Lunar mission}

KAGUYA (SELENE) was launched on September 14, 2007 by H-IIA to study the structure and the origin of the Moon. NAOJ is responsible for topography measurement using a laser altimeter and gravity measurement using relay and VLBI subsatellites.  NAOJ team reports lunar global shape and topography with two order of magnitude higher density and better vertical resolution than previously available. The result shows that the highest point on the Moon is on the southern rim of the Dirichlet-Jackson basin, and the lowest one is in the Antoniadi crater. The full-range topography spans about 19.81km, which is greater than previous estimation \citep{araki09}. Detailed maps of polar regions were obtained for the first time (cf.Fig.14).  A dramatically improved lunar gravity model is also reported \citep{namiki09}. The model includes far side tracking data for the first time. The results reveal that the far side impact basins do not resemble their nearside counterparts. The researchers interpret the results as indicating cooler, more rigid early conditions on the far side than previously thought. VLBI observations using VERA and international stations contribute significant improvement of orbital determination of satellites, which would improve lunar gravity model.

\begin{figure}[htbp]
\label{kaguya}
\begin{center}
\includegraphics[width=8cm,clip]{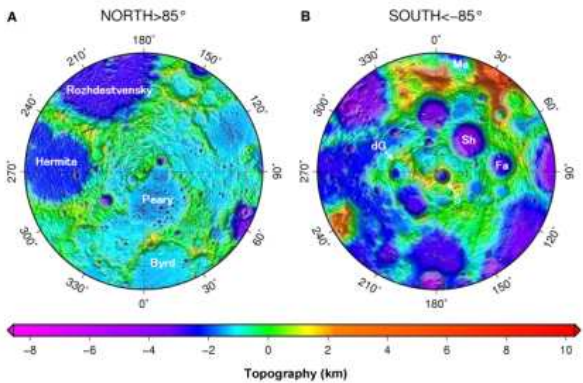}
\caption{Maps of Lunar northern polar region (left) and southern polar region (right) first obtained by Kaguya mission.
}
\end{center}
\end{figure}

\section{Centers and Divisions}
Advanced Technology Center(ATC) has machine shop, optical shop, space chamber shop (Fig. 15), design shop, high precision machining facilities and a coating facility. It was a center for developing Subaru Telescope instruments in late 90's. Currently, active projects run at ATC include band 4, 8, and 10 receiver developments for ALMA, R\&D efforts for tera hertz detectors, gravitational wave detector and light weight mirror technology.

\begin{figure}[htbp]
\label{clean}
\begin{center}
\includegraphics[width=7cm,clip]{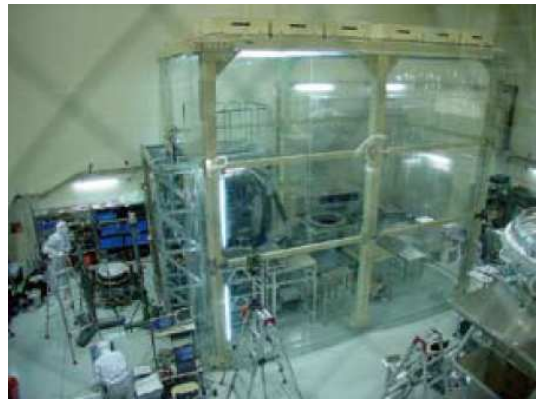}
\caption{Large clean room for testing the performances of telescope and instruments for space missions.
}
\end{center}
\end{figure}

Astronomy Data Center (ADC) offers computer systems for astronomical data analysis for astronomers and operates astronomical data archives center. SMOKA provides access to the public science data obtained at Subaru Telescope, 188cm telescope at Okayama Astrophysical Observatory, and 105cm Schmidt telescope at Kiso Observatory (University of Tokyo). Japanese Virtual Observatory project is also a part of ADC's activity.

Center for Computational Astrophysics (CfCA) has several supercomputers dedicated to astrophysical large-scale simulations. Currently, CfCA hosts NEC SX-9 vector supercomputer (2Tflops) and Cray XT4 parallel supercomputer (28Tflops), as well as GRAPE-6 and 7 special-purpose computers for gravitational N-body simulations. These facilities are used by astrophysicists both within and outside Japan, for studies of supernova explosion, accretion disks, star formation, planet formation, galaxy formation, cosmology, and other areas of theoretical astrophysics. Fig. 16 shows the result of recent self-consistent simulation of a disk galaxy performed on the Cray system, in which gas cooling is followed down to 10K, resulting in the spiral arms which look quite similar to that of real spiral galaxies.

Public Relations Center (PRC) provide dedicated services for public outreach, publications, library, and ephemeris computation.

\begin{figure}[htbp]
\label{baba}
\begin{center}
\includegraphics[width=7cm,clip]{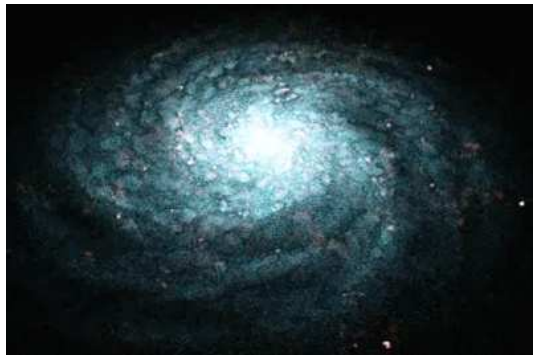}
\caption{A spiral galaxy as reproduced in a self-consistent simulation incorporating physica processes.
}
\end{center}
\end{figure}

There are four divisions for Optical and Infrared Astronomy, Radio Astronomy, Solar and Plasma Astrophysics, and Theoretical Astronomy, where many of individual researchers are affiliated.

\section{Other Space Missions}
As for the Hinode mission, a new solar observatory in space, see \citep{tsun09}. NAOJ astronomers are collaborating with JAXA scientists for research and development acitivities to realize 3.5m SPICA(SPace Infrared Telescope for Cosmology and Astrophysics) mission, a mid- and far-infrared astronomy mission with a cooled (4.5 K) telescope launched into Sun-Earth L2 orbit, VSOP-2, a space VLBI mission following the VSOP (Haruka) to image AGNs at 30 micro arcsec resolution, and JASMINE(Japan Astrometry Satellite Mission for INfrared Exploration) to measure position, distance and proper potions of stars in the bulge of the Galaxy. 

\vspace{5mm}
Upon an urgent request from the editor, the present paper was written by drafting the outline of the NAOJ and compiling the descriptions available in recent NAOJ web releases and those given in NAOJ 2008 annual report \citep{naoj08} with some editing and revisions. The author acknowledges helpful quick comments delivered by M.fujimoto, R.Kawabe, J.Makino, T.Sakurai, S.Sasaki, Y.Suematsu, K.Tatematsu, S.Tsuneta, and M.Yoshida to improve the first draft.

\end{document}